# Smart Temperature Sensor for Thermal Testing of Cell-Based ICs

S.A. Bota, M. Rosales, J. L. Roselló, J. Segura

Grup de Tecnologia Electrbnica. Universitat de les Illes Balears. Campus UIB. 07122 Palma. Spain.
email sebastia.bota@uib.es

**Abstract**

*In this paper we present a simple and efficient built-in temperature sensor for thermal monitoring of standard-cell based VLSI circuits. The proposed smart temperature sensor uses a ring-oscillator composed of complex gates instead of inverters to optimize their linearity. Simulation results from a 0.18-μm CMOS technology show that the non-linearity error of the sensor can be reduced when an adequate set of standard logic gates is selected.*

## 1. Introduction

The increase of circuit density and clock speed produce an increase in power consumption that have brought thermal issues into the spotlight of ULSI design. It has been reported that the measured junction temperature of a 1-GHz 64-bit RISC microprocessor implemented in 0.18-um CMOS technology is as high as 135 °C at $V_{DD}$=1.9V [1], technology scaling makes this trend more severe and the junction temperature of a 0.13-um CMOS chip has estimated to be 3.2 times higher than the junction temperature of 0.35-um CMOS chip [2] working under equivalent conditions.

It is well known that high temperature operation compromises long-term IC reliability and impacts circuit performance. To avoid system failure, design techniques for thermal testability and thermal management have been incorporated into several electronic products [3].

Temperature sensors are the core part of any thermal management system. Because design methodologies for huge digital systems are driven by high-level functional descriptions, the use of temperature sensors based on classical analogue circuit topologies [4] seems to be inadequate in a large number of applications. Their demand of full-custom design steps and the need to convert temperature (related to an analogue voltage or current signal) to a digital magnitude are the major drawbacks. In this paper we have investigated how to implement a thermal monitoring system suitable for deep-submicron cell-based designs.

## 2. Ring-oscillator temperature sensors

Diode-based sensors are used as temperature detection mechanism in Pentium 4 processors [4] or in the Thermal Assist Unit of the PowerPc processors [3]. Unfortunately, in cell-based design styles (where only digital gates are available) this kind of sensors result difficult to implement.

It is well-known that the delay of a logic gate increases with temperature, therefore a way to measure the junction temperature of silicon is to use a ring-oscillator. In [5] an application based on a ring oscillator implemented in FPGAs was analyzed. Taking into account that both Standard-cells and FPGA are cell-based design styles, the translation of the FPGA design to standard-cells is straightforward, anyway, it will be interesting to exploit the higher flexibility related to the standard-cell style to improve sensor performances in terms of area, sensor calibration, and linearity.

A ring-oscillator consists of an odd number of inverters connected in a circular chain [6]. The circuit oscillates with a period equal to:

$$T_{osc} = N\left(t_{pHL} + t_{pLH}\right) \qquad (1)$$

with N the number of inverter stages in the chain, $t_{pHL}$ the high-to-low transition delay and $t_{pLH}$ the low-to-high transition delay of the composing gates. The simulated response of a ring oscillator is shown in Fig. 1.

The behavior of $T_{osc}$ has been computed for different channel width ratios of N and P transistors ($W_p/W_n$) in a 5-inverter ring-oscillator. According to the results shown in Fig. 2, by optimizing the circuit at transistor level, it is possible to reduce the non-linearity error in the range of temperatures of interest (-50°C to 150°C) below 0.2%.

This work has been partially supported under the Spanish Government project CICYT-TIC0201238 and CRL-Intel Research Laboratories.

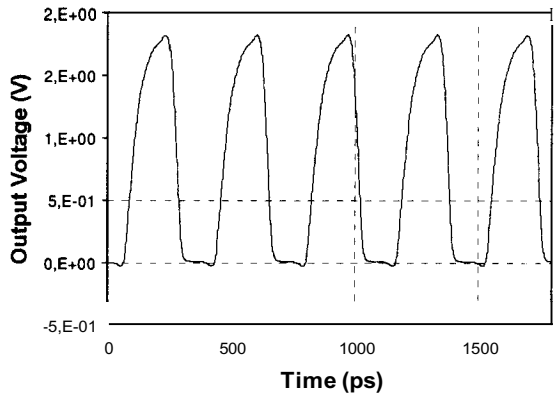

Fig. 1. HSPICE simulation of a ring oscillator output of a five-stage inverters.

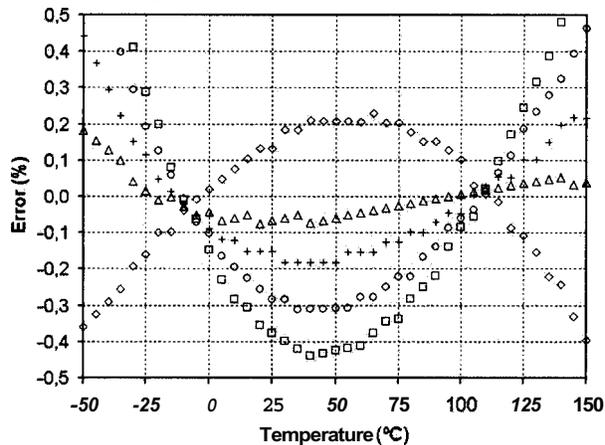

Fig. 2. Non-linearity error for different Wp/Wn ratios. (◊) Wp/Wn=1, (△) 1.75, (+) 2.25, (○) 3, (□) 4.

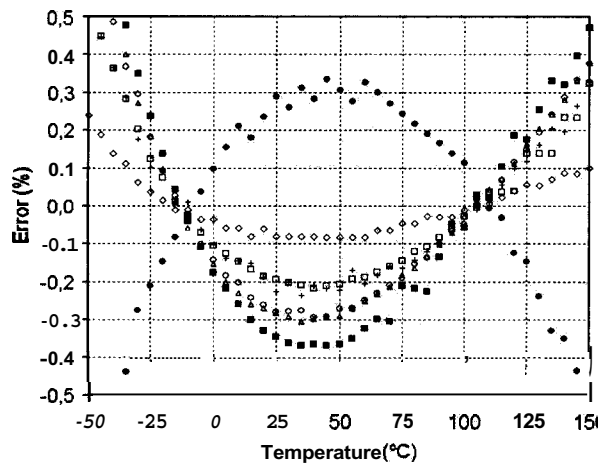

Fig. 3. Non-linearity error for different ringoscillator configurations: (●) 5 NOR2, (◊) 3 INV + 2 NOR2, (□) 3 NAND3 + 2 NOR2, (+) 5 INV, (○) 2 INV + 3 NAND2, (△) 5 NAND2 and (■) 2 INV + 3 NAND4.

A low dependence with the number of inverting stages has been found, ring-oscillators with **5,** 9 or 21 stages have similar characteristics in terms of linearity.

As a summary, the results of this section suggest that the ring-oscillator must be optimized at the transistor level; unfortunately these sizes can be different to those used in the inverters of the target standard-cell library. If an smart temperature sensor has to be included in a given cell-based digital system, the way to avoid any full-custom step in the design process consist in designing the proposed ring-oscillator only using standard library cells.

## 3. Cell-Based Optimization

The linear dependence of propagation delay with temperature can also be adjusted replacing the inverters by other inverting logic gates, like **NAND** and **NOR** gates. This introduces a new degree of freedom in the design. In Fig. 3 we present the non-linearity error for different cell configurations. With the proposed method, the error of the ring-oscillator can be reduced since a 0.2%, similar to the error obtained when changing the transistor sizes.

An smart unit for thermal management have been designed using the ring-oscillator as a temperature sensor and an additional digital processing bloc to convert oscillation period to temperature expressed in digital format. The possibility to disable the oscillator in order to minimize self-heating, produce an output signal to indicate that a measurement is in progress or multiplexing the readout from different ring-oscillators distributed on different points for thermal mapping are examples of other features of this unit.